\def\year{2020}\relax
\documentclass[letterpaper]{article} 
\usepackage{aaai20}  
\usepackage{times}  
\usepackage{helvet} 
\usepackage{courier}  
\usepackage[hyphens]{url}  
\usepackage{graphicx} 
\usepackage{graphicx}  
\usepackage{subcaption}
\usepackage{comment}
\usepackage{url}
\usepackage{verbatim}

\usepackage[utf8]{inputenc}
\usepackage[inline]{enumitem}
\usepackage{booktabs}
\usepackage{amssymb}
\usepackage{amsmath}
\usepackage{multirow}
\usepackage{latexsym}
\usepackage{soul}
\usepackage{array}

\title{Charting the Landscape of Online Cryptocurrency Manipulation}
\author{Leonardo Nizzoli,\textsuperscript{\rm 1} Serena Tardelli,\textsuperscript{\rm 1} Marco Avvenuti,\textsuperscript{\rm 2} Stefano Cresci,\textsuperscript{\rm 3} Maurizio Tesconi,\textsuperscript{\rm 3} Emilio Ferrara\textsuperscript{\rm 4}\\ %
\textsuperscript{\rm 1} Dept. of Information Engineering, University of Pisa and IIT-CNR, Italy \ \ [name.surname]@iit.cnr.it \\
\textsuperscript{\rm 2} Dept. of Information Engineering, University of Pisa, Italy \ \ marco.avvenuti@unipi.it\\
\textsuperscript{\rm 3} Institute of Informatics and Telematics, IIT-CNR. Pisa, Italy \ \ [name.surname]@iit.cnr.it\\
\textsuperscript{\rm 4} Information Sciences Institute, University of Southern California. Marina del Rey, CA (USA) \ \ ferrarae@isi.edu}

\begin{document}

\maketitle

\begin{abstract}
Cryptocurrencies represent one of the most attractive markets for financial speculation. As a consequence, they have attracted  unprecedented attention on social media. Besides genuine discussions and legitimate investment initiatives, several deceptive activities have flourished. In this work, we chart the online cryptocurrency landscape across multiple platforms. To reach our goal, we collected a large dataset, composed of more than 50M messages published by almost 7M users on Twitter, Telegram and Discord, over three months. We performed bot detection on Twitter accounts sharing invite links to Telegram and Discord channels, and we discovered that more than 56\% of them were bots or suspended accounts. Then, we applied topic modeling techniques to Telegram and Discord messages, unveiling two different deception schemes  -- ``pump-and-dump'' and ``Ponzi'' -- and identifying the channels involved in these frauds. Whereas on Discord we found a negligible level of deception, on Telegram we retrieved 296 channels involved in pump-and-dump and 432 involved in Ponzi schemes, accounting for a striking 20\% of the total.  Moreover, we observed that 93\% of the invite links shared by Twitter bots point to Telegram pump-and-dump channels, shedding light on a little-known social bot activity. Charting the landscape of online cryptocurrency manipulation can inform actionable policies to fight such abuse.
\end{abstract}


\section{Introduction} \label{sec:introduction}
The explosive growth of online social media has changed the way we interact, cooperate, make money and get information about diverse topics. 
Social media have also grown in popularity as a fascinating online showcase in which to promote and help the cryptocurrency world to grow~\cite{thelwall2018can}. As cryptocurrency is thriving worldwide, financial institutions and countries are debating about its eligibility as a payment system. However, there are no explicit laws and regulations on cryptocurrency markets by any government or institution yet, as the main peculiarities of cryptocurrencies are decentralization, highly speculative nature and self-regulation~\cite{krafft2018experimental}.
For this reason, the cryptocurrency domain lends itself as an excellent breeding ground for manipulation to run wild. Such manipulations are often enabled with social media. In fact, online ecosystems are a suitable habitat for deception, which can be achieved with minimal efforts and may result in
high success rate~\cite{tsikerdekis2014online}.
Therefore, there is a growing need to understand how susceptible cryptocurrency markets are to online manipulation.

Literature has investigated forms of fraud such as cryptocurrency thefts~\cite{chohan2018problems}, ``Ponzi''~\cite{chen2018detecting} and ``pump-and-dump'' schemes~\cite{xu2019anatomy}, mainly by focusing on discussion forums and market data. Instead, the role of social media in promoting these frauds has been largely overlooked and is still unclear. Regarding frauds, Ponzi scheme is a financial scam that relies on acquiring investors by promising high returns in exchange of a minimum amount of currency. Those funds are used to generate profits for old investors and organizers. When the rate of new investors is not large enough to sustain the process, the chain breaks, and last comers lose their investment. 
Another well-known type of scam is the pump-and-dump scheme, where participants collectively aim to artificially inflate a currency price through coordinated, simultaneous buying (``pump''). Once outside unaware investors notice the surge in price and start investing in the asset, the participants sell to them, thus making a profit and causing a price collapse (``dump''). Generally, there are orchestrators behind the curtains, who profit even at the expense of the witting participants themselves, let alone the other unaware investors. Social media platforms, such as Telegram and Discord, are the perfect habitat for such scams to proliferate. They offer anonymity and low levels of moderation. Moreover, they feature channels as a way to broadcast public messages to large audiences and to invite investors to join. Indeed, most successful scams depend on attracting a large mass of users.

This study moves in the direction of investigating a multi-platform social ecosystem in which cryptocurrencies are discussed, in order to uncover possible deceptive schemes and to assess their extent. We pose our focus on the diffusion of invite links -- that is, special URLs allowing users to join channels. In fact, our controlling idea for this study is based on the intuition that fraudsters can exploit invite links to scam channels as an effective way for recruiting participants to the scam. In addition, diffusion patterns of invite links already provided valuable information for detecting homophily and common interests in online communities~\cite{anderson2015global}. In detail, we first collect a rich dataset of 16M tweets posted between March and May 2019, discussing 3K cryptocurrencies. Then, starting from those tweets, we retrieve invite links to Telegram and Discord channels. We then crawl channel messages, looking for new invite links, and we repeat the process in an iterative, snowball fashion. We end up with an unprecedented multi-platform dataset, composed of 10M Discord and 23M Telegram messages, in addition to the 16M tweets. We emphasize the interplay between agents within and among different platforms, 
by enriching our initial dataset via an analysis of the nature (i.e., deceptive \textit{vs} legitimate) of Twitter accounts and via content analyses for Telegram and Discord channels. This choice takes into account the differences between these platforms, affecting how deception is implemented.
Firstly, we focus on the genuineness of Twitter accounts involved in the discussion, looking for traces of deceptive behaviors. We discover a significant presence of suspended accounts (19.9\%) and bots (36.4\%), labeled by a state-of-the-art technique.
Secondly, we perform topic modeling on Telegram and Discord channels, in order to label each channel according to its dominant topic. Two types of cryptocurrency manipulations emerge, resulting in 297 channels performing pump-and-dump and 432 channels performing Ponzi schemes. Only one pump-and-dump channel is hosted on Discord, while all the other deceptive channels belong to Telegram. Accordingly, Discord emerges as a reasonably healthy ecosystem contrarily to Telegram, where 56.5\% of the cryptocurrency-related channels of our dataset are involved in deception. 
At last, we cross-check these different types of manipulation. We find out that deceptive channels are the ones receiving the most part of the invite links (87.8\%), confirming the founding intuition of our controlling idea. Moreover, pump-and-dump channels collect 92.9\% of the invite links broadcast by Twitter bots. 
This is evidence that Twitter social bots promote deceptive cryptocurrency content, created for the explicit purpose of spamming invite links to pump-and-dump channels.
This study strongly contributes to mapping the role of automated accounts in spreading misinformation through social media, and it is a good starting point to measure the effects of social media disinformation and misinformation in the real world.


\noindent\textbf{Contributions of this work.}
We summarize our main contributions in the following:

\begin{itemize}
    \item We collect and share a large dataset for studying online cryptocurrency manipulations, comprising more than 50M messages and describing the online cryptocurrency ecosystem across three major platforms -- such as Twitter, Telegram and Discord.
    \item We uncover the pivotal role of Twitter bots in broadcasting invite links to deceptive Telegram and Discord channels, exposing a little-known social bot activity.
    \item Instead of focusing on specific frauds, we let manipulation patterns naturally emerge from data, highlighting the existence of 2 different manipulations -- namely, \textit{pump-and-dump} and \textit{Ponzi} schemes. 
    \item Our results ultimately demonstrate that Discord can be considered as a reasonably healthy online cryptocurrency ecosystem. In contrast, more than 56\% of all Telegram channels are involved in manipulations. Moreover, these deceptive activities are massively broadcast with the help of Twitter bots.
\end{itemize}

\noindent\textbf{Reproducibility.} We will release an anonymized, privacy-preserving version of the dataset upon the paper acceptance.




\section{Related works}
\label{sec:related}

\noindent\textbf{Cryptocurrency manipulations.} There is still limited literature on the analysis of cryptocurrencies involving social media. The majority of previous studies focused on predicting the effects of specific manipulation activities in cryptocurrency markets. As an example,~\cite{xu2019anatomy} provided a detailed description of pump-and-dump schemes on Telegram, and developed a model to predict the likelihood of a coin being the target for manipulation. More akin to our work is~\cite{mirtaheri2019identifying}, where authors collected pump-and-dump Telegram channels in a snowball fashion, starting from a seed of known pump-and-dump Telegram channels. They focused on predicting the presence and success of pump operations in terms of meeting the anticipated price targets. They did so by leveraging market and Twitter data, and they observed the presence of suspicious Twitter users. 
Other studies observed that pump-and-dump phenomena are widespread on both
Discord and Telegram~\cite{hamrick2018economics}, and they evaluated the effects of such activities on the liquidity and price of cryptocurrencies~\cite{li2018cryptocurrency}.
Pump-and-dump is not the only financial fraud under scrutiny. In~\cite{vasek2018analyzing}, authors investigated online Ponzi schemes, advertised on the \textit{Bitcointalk} discussion forum. They used survival analysis to identify factors that affect scam persistence. In~\cite{bartoletti2018data}, authors proposed machine learning algorithms for automatic classification of Ponzi schemes involving cryptocurrencies. These important studies have offered key insights into specific cryptocurrency manipulation domains. Instead, here our goal is to offer a more general insight into the bounds of cryptocurrency manipulation activities that take place across multiple social media platforms. 
Widening the perspective beyond social media, other works focused on discussion forums to identify and characterize communities by applying topic modeling~\cite{linton2017dynamic}, or to predict fluctuation in cryptocurrency prices~\cite{kim2016predicting}.
In~\cite{krafft2018experimental}, authors tried to determine the dynamics of cryptocurrency markets, and demonstrated that trading bots can alter market behavior. 

\noindent\textbf{Other online manipulations.} The existence of manipulative, deceptive, synthetic content in online discussions has already been witnessed in a wide variety of societal topics. 
For instance, it has been demonstrated that bots are exploited to promote online financial content~\cite{cresci2019cashtag}, as well as health content~\cite{allem2018could}. Other studies showed that bots tampered with US~\cite{bessi2016social,addawood2019linguistic}, Japanese~\cite{schafer2017japan}, South Korean~\cite{keller2017manipulate}, French~\cite{ferrara2017disinformation}, Italian~\cite{cresci2017tdsc}, and German~\cite{Kupferschmidt1081} political elections. 

In other recent work~\cite{ferrara2016rise}, it is reported the emergence of new waves of social bots, capable of mimicking human behavior in social media better than ever before. As social bots evolve, online content manipulation goes undetected even by platform administrators~\cite{cresci2017paradigm}, with consequent profound impact on content popularity and activity in social media~\cite{aiello2012people}. Scholars and platform administrators reacted by proposing more advanced detection techniques based on the analysis of both individual~\cite{davis2016botornot} and collective~\cite{cresci2017paradigm} behaviors. The current research trend with regards to online manipulation is shifting from a focus on individual malicious accounts (e.g., bots, trolls) to a broader and more sophisticated model that embraces the interplay between both automated and human-driven behaviors~\cite{starbird2019disinformation}. However, to the best of our knowledge, the latter model is yet to be exploited and operationalized.

\section{Designing and collecting the dataset}
\label{sec:dataset}

\begin{table}[t]
\centering
\scriptsize
\begin{tabular}{lrrr}
\toprule
& \textbf{channels} & \textbf{users} & \textbf{messages} \\
\midrule
Discord              & 1,755                                                                       & 211,409                                                                  & 10,331,720                                                                  \\
Telegram             & 3,813                                                                       & 920,925                                                                  & 23,812,537                                                                  \\
Twitter              & --                                                       & 5,745,944                                                                & 16,840,312                                                                  \\ \midrule
total                & 5,568                                                                       & 6,878,278                                                                & 50,984,569                                                                  \\ \bottomrule
\end{tabular}
\caption{Counts of distinct channels, users and messages for each considered platform.} 
\label{tab:dataset_counts}
\end{table}

\noindent\textbf{Preliminaries.} Previous works about cryptocurrency manipulation~\cite{xu2019anatomy,vasek2018analyzing} focused on a specific scheme (e.g., pump-and-dump or Ponzi), aiming to outline its anatomy, assess its efficacy or predict its occurrence. Accordingly, they relied on datasets specifically designed to include only data pertinent to the cryptocurrency manipulation scheme under exam. Conversely, here we are interested in performing an unprecedentedly wide, horizontal exploration of the online cryptocurrency ecosystem, including multiple platforms and avoiding any bias towards legitimate or deceptive communities. In this way, we have the chance to 
\begin{enumerate*}[label=(\roman*)]
  \item observe deceptive schemes naturally emerging from the data,
  \item assess their spread within the online cryptocurrency ecosystem, 
  \item identify legitimate and deceptive agents (e.g., accounts, channels), and
  \item study the interplay between them.
\end{enumerate*}
In order to obtain a dataset with the desired features, we designed and implemented a crawling strategy based on a snowball approach. We focused on the Twitter microblogging platform, and on two instant messaging platforms known to host cryptocurrency communities: Telegram and Discord.

Telegram features two types of group chats: 
\begin{enumerate*}[label=(\roman*)]
  \item \textit{groups} -- where all members have privilege to share contents by default, and
  \item \textit{channels} -- where usually only administrators broadcast contents to their audience. 
\end{enumerate*}
They can be joined by means of specific invite links (URLs), which can contain the required password in case the group or channel is private. Discord features \textit{servers} (also referred to as \textit{guilds}), in which admins can create several channels -- each one usually devoted to a specific topic -- and handle the writing privileges. Authorized users can generate invite links (URLs) for the server, which are specific for the user who created them. Hereafter, we use the generic term ``channel'' for Telegram groups and channels as well as for Discord servers, and the term ``invite link'' for every type of URL allowing users to join a channel.

\begin{figure}[t]
    \centering
    \begin{subfigure}[t]{.4\linewidth}
        \includegraphics[width=\linewidth]{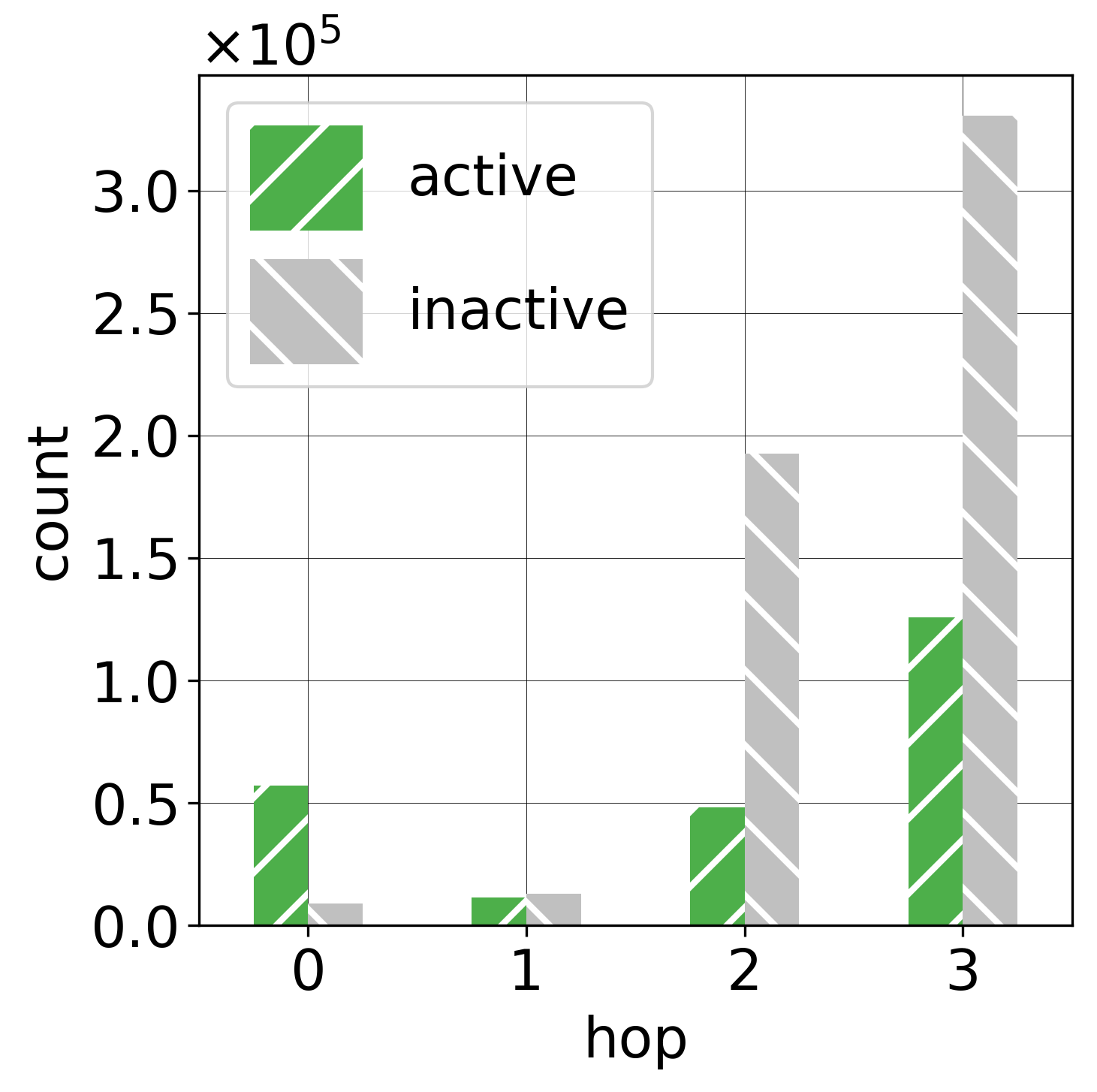}
        \caption{Total invite links (Telegram).}
        \label{fig:telegram_tot_invites}
    \end{subfigure}\hspace{0.03\textwidth}%
    \begin{subfigure}[t]{.4\linewidth}
        \includegraphics[width=\linewidth]{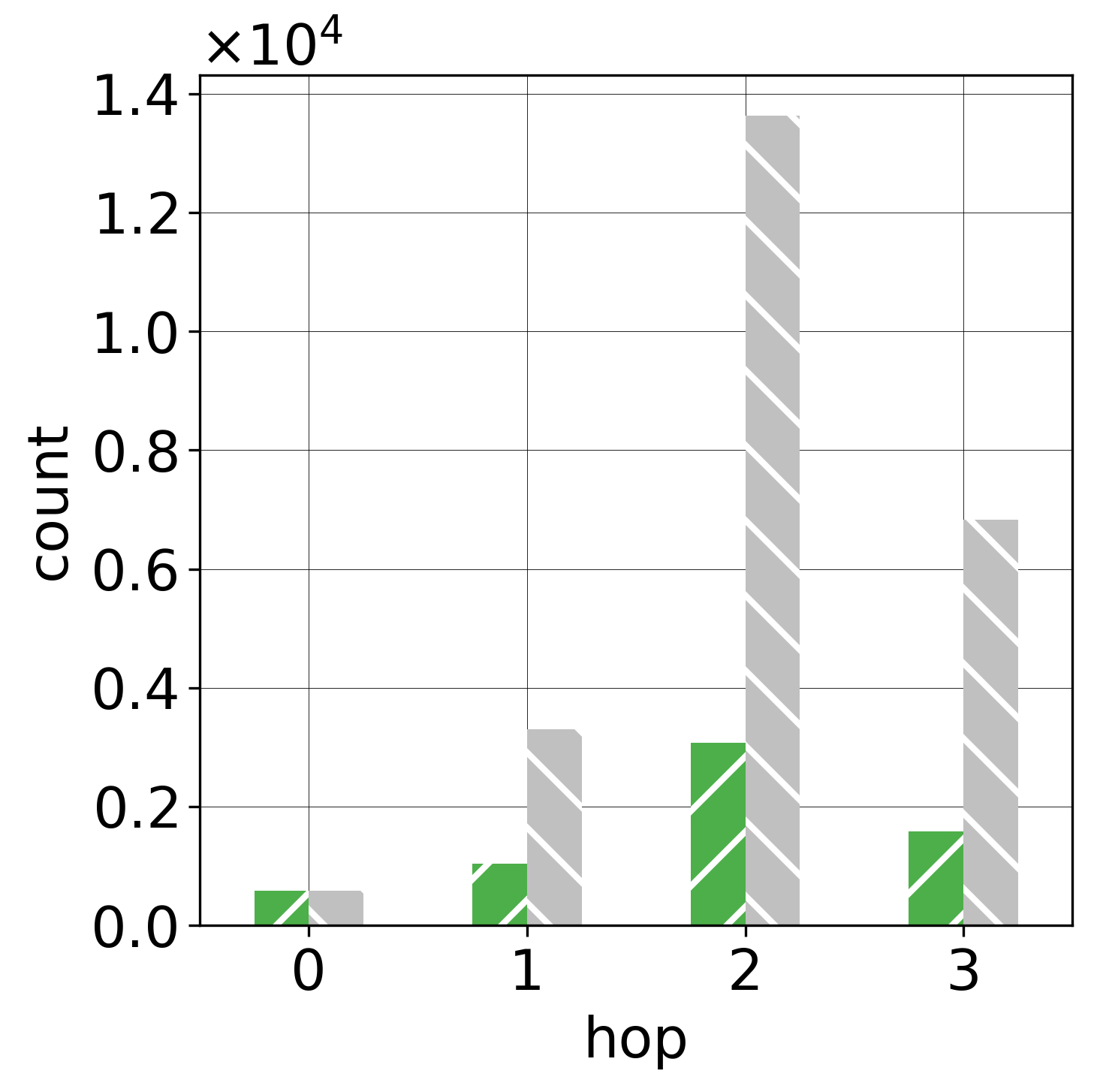}
        \caption{Distinct invite links (Telegram).}
        \label{fig:telegram_dist_invites}
    \end{subfigure}\hspace{0.03\textwidth}%
    \begin{subfigure}[t]{.4\linewidth}
        \includegraphics[width=\linewidth]{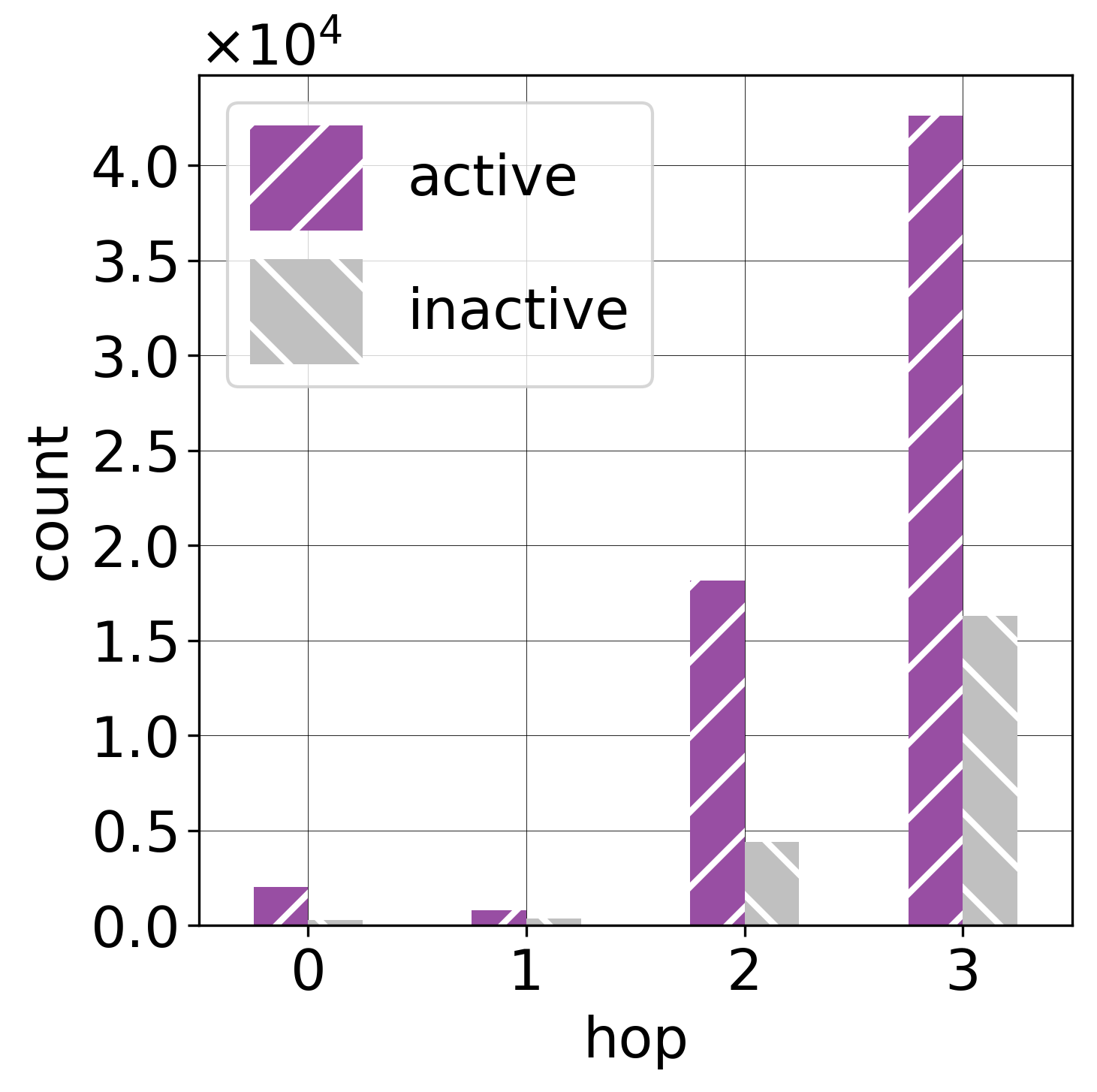}
        \caption{Total invite links (Discord).}
        \label{fig:discord_tot_invites}
    \end{subfigure}\hspace{0.03\textwidth}%
    \begin{subfigure}[t]{.4\linewidth}
        \includegraphics[width=\linewidth]{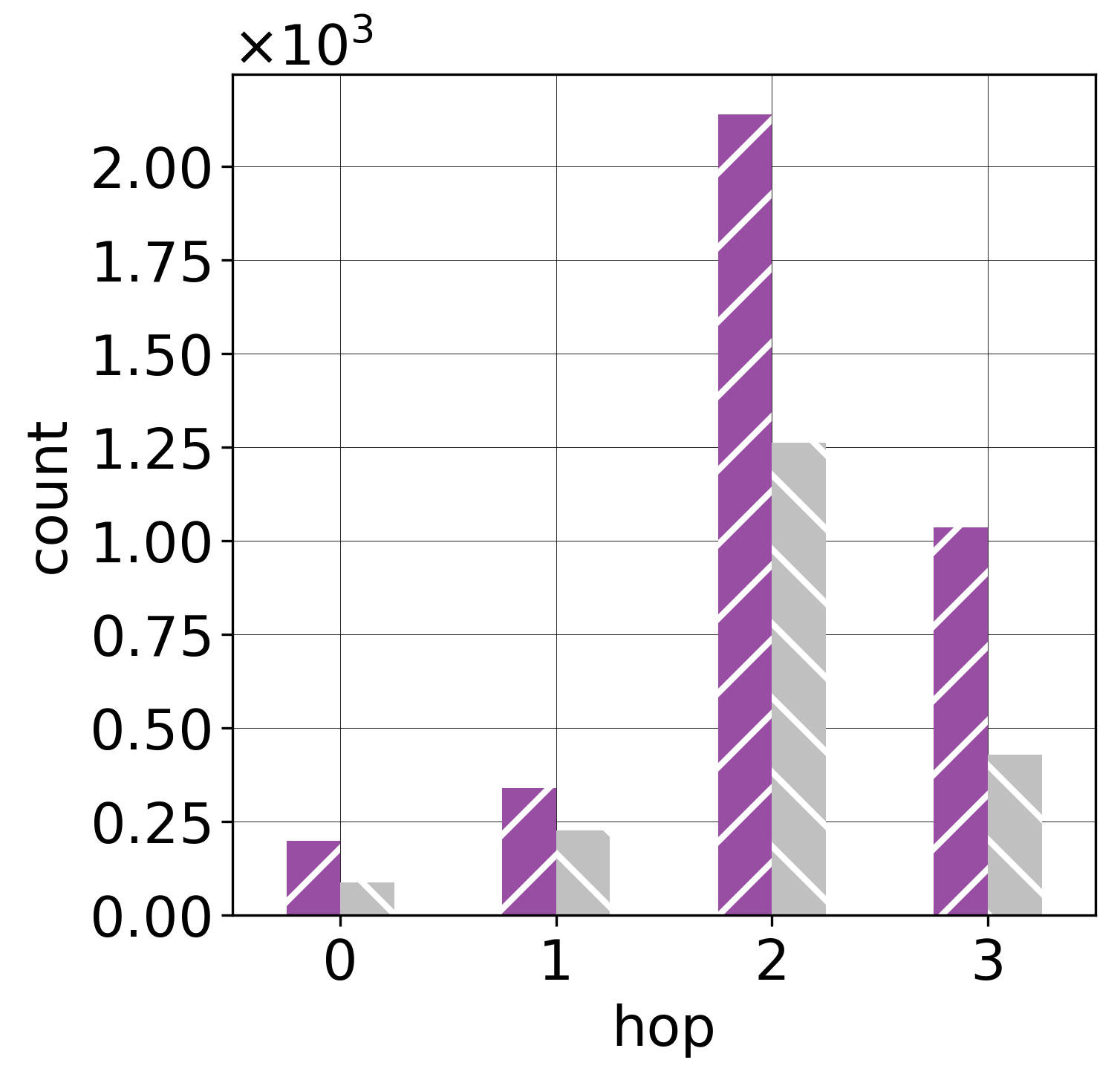}
        \caption{Distinct invite links (Discord).}
        \label{fig:discord_dist_invites}
    \end{subfigure}
    \caption{
    Counts of active and inactive invite links to Telegram (1a, 1b) and Discord (1c, 1d) channels, retrieved at each hop of our snowball crawling strategy.
    Telegram attracts much more invite links than Discord (79.2\%). The large number of inactive invite links may reflect the practice of publishing “expiring” invite links.
    }
    \label{fig:invites_hop}
\end{figure}

\noindent\textbf{Data collection.} Firstly, we leveraged Twitter's Streaming API to collect all tweets mentioning at least one of the 3,822 cryptocurrency cashtags\footnote{The cashtag of a cryptocurrency is composed of a dollar sign followed by the ticker symbol of the cryptocurrency (e.g., \$BTC for Bitcoin). Similarly to hashtags, they can be used to efficiently tag and filter tweets.} provided by the \textit{CryptoCompare}\footnote{\url{https://www.cryptocompare.com/}} public API. This data collection covered a three months-long time window spanning from March to May 2019, and resulted in the acquisition of more than 16M tweets. Then, we retrieved all the invite links contained in these tweets pointing to Telegram or Discord channels and we used them as seeds for an iterative snowball crawling strategy. In particular, this first set of channels, pointed by those invite links, represents the \textit{hop 0} of our crawl. By leveraging Telegram and Discord APIs, we collected the message histories of hop 0 channels. Then, we parsed such messages looking for more invite links; we retrieved the message histories of the related (hop 1) channels, and we continued iterating this data collection pipeline. At hop 3, we retained only invite links pointing to channels already found at hops 0-2, and we concluded our crawling.
In Table~\ref{tab:dataset_counts}, we provide some aggregates of the obtained dataset. As shown, our dataset includes more than 50M messages, published by almost 7M distinct users across the three platforms. In particular, we highlight the unprecedentedly large number of Telegram (3,813) and Discord (1,755) channels, that guarantees a sound coverage of the cryptocurrency ecosystem on such platforms. Focusing on the two instant messaging platforms, we notice that 68.5\% of the retrieved channels, 81.3\% of distinct users and 69.7\% of messages belong to Telegram. In Figure~\ref{fig:invites_hop}, we depict the count of active and inactive invite links retrieved at each hop of our snowball crawling strategy. Considering the combined amount of invite links for both platforms, Telegram accounts for 79.2\% of active and 96.2\% of inactive links.
We highlight that our data collection strategy is impartial with respect to the two instant messaging platforms. Hence, we can conclude that Telegram is much more used than Discord within the online cryptocurrency ecosystem. Finally, figures~\ref{fig:telegram_tot_invites}, \ref{fig:telegram_dist_invites} show that at hop $>1$ inactive Telegram invite links largely exceed active ones, as opposed to Discord (figures~\ref{fig:discord_tot_invites}, \ref{fig:discord_dist_invites}). The large number of inactive invites may reflect the practice of publishing ``expiring'' links, to promote more elitist, limited access channels.

As an additional contribution of our work, we publish an anonymi\-zed, privacy-preserving version of this dataset,\footnote{The dataset URL will be included upon paper acceptance.} to allow the reproducibility of our experiments and foster further research on this important topic.

\section{Building the invite link network}\label{sec:network}

\begin{figure}[t]
    \centering
    \includegraphics[width=0.7\linewidth, angle=0]{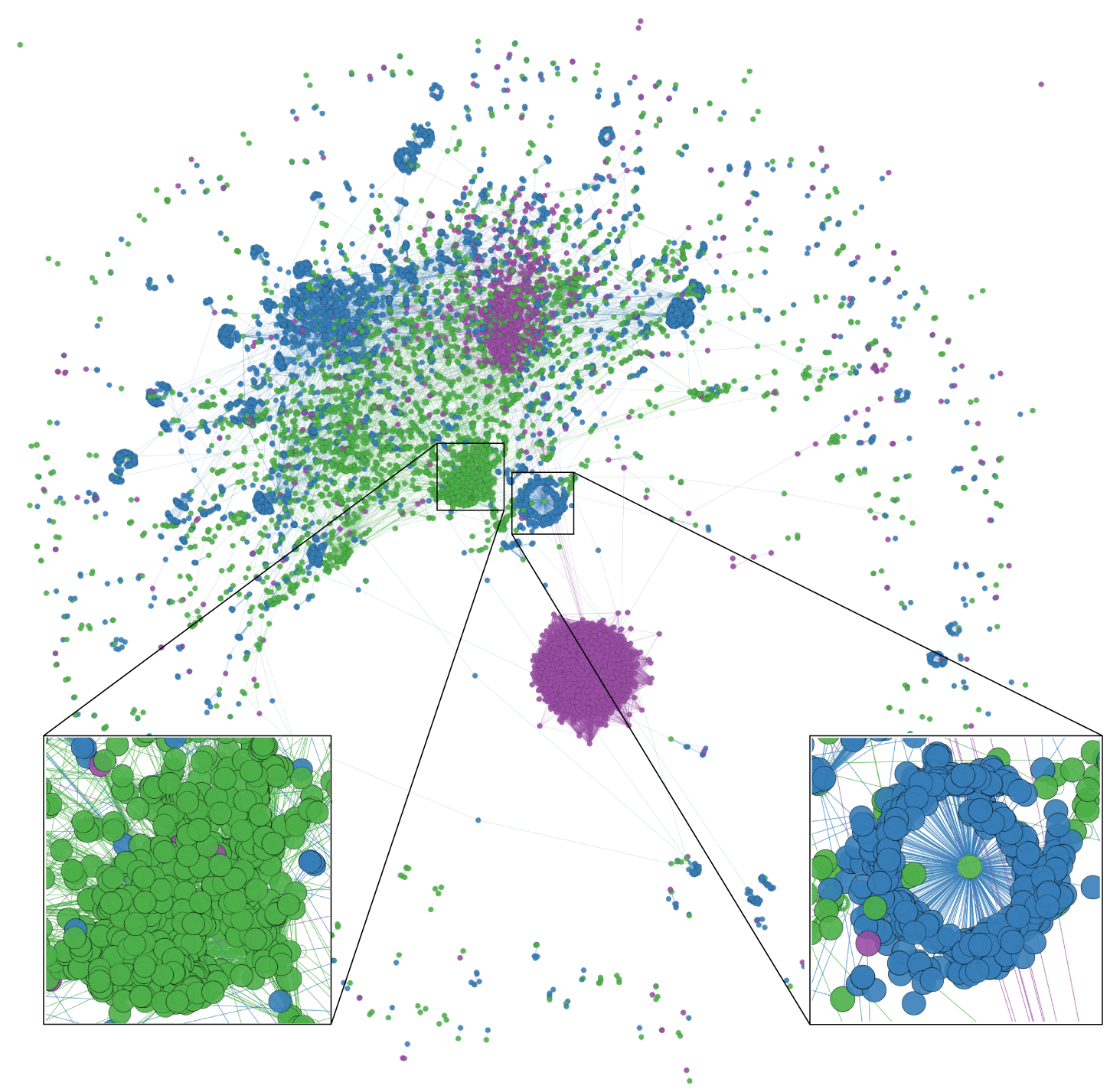}
    \caption{ForceAtlas node-link diagram of the invite link network. Nodes represent Twitter accounts (blue), Telegram (green) and Discord channels (violet) sharing or receiving at least one invite link. Edges are colored by their source node color. Peculiar network structures emerge, such as a dense cluster of Telegram channels (bottom-left inset) and star structures (bottom-right inset).}
    \label{fig:network-graph}
\end{figure}

The diffusion of invite links plays a major role in the growth of online platforms and communities. Moreover, there is a strong interplay between the structural properties underlying the diffusion of invites and the characteristic features of the source and target agents involved in those processes~\cite{anderson2015global}. In particular, the exchange of invites can be an excellent proxy for homophily or common goals. Moreover, the effectiveness of cryptocurrency manipulation schemes depend on the number of participants involved. Hence, we hypothesize deceptive agents to give a major contribution to the diffusion of invite links. For this reason, we study this process by building the invite link network, which is shown in Figure~\ref{fig:network-graph}. It is a directed, weighted network composed of 13,009 nodes and 62,278 edges. Nodes represent agents sharing or receiving at least one invite link. In detail, 7,441 (57.2\%) nodes are Twitter accounts, 3,813 (29.3\%) are Telegram channels and 1,755 (13.5\%) are Discord channels. Edges are directed from a source node -- representing an agent who broadcasts an invite link, to a target node -- representing the channel pointed by the invite link, and their weights account for the number of existing invite links between the two. It is worth noticing that Twitter nodes can only have outgoing edges, since Twitter accounts cannot receive invite links.

Figure~\ref{fig:network-graph} shows a node-link diagram  of the network, realized with the \textit{ForceAtlas} algorithm. Node size is proportional to the number of members of a channel or the number of followers of a Twitter account -- that is, to the size of the potential audience of the agent. Nodes are colored according to the corresponding platform. Edge thickness is proportional to the weight, and their color is the same as that of the source node. ForceAtlas determines the layout of nodes so that nodes connected by strong links appear close to each other in the diagram. Figure~\ref{fig:network-graph} shows the presence of a giant component, including 91\% of nodes and having a diameter of 19. Since the giant component includes most of the nodes and links, we focus the rest of our analyses on it. The giant component includes a strongly clustered community of Discord channels, weakly connected to the rest of the nodes and represented as an isolated violet ``hairball'', located near the center of Figure~\ref{fig:network-graph}. Within the rest of the giant component, there is still a clear separation between Discord and Telegram nodes. A very dense cluster of Telegram channels (green-colored) is magnified in the bottom-left corner of the plot. Twitter nodes (blue-colored) appear as frequently arranged in a ring surrounding a single channel -- usually a Telegram one -- thus forming a star structure. An example of this feature is magnified in the bottom-right corner of the plot. We counted 135 of these structures having a size of at least 10 accounts, 107 of which (79.2\%) are centered on a Telegram channel. Interestingly, these preliminary results highlight the presence of peculiar network structures (e.g., dense clusters, stars), likely representative of some interesting real-world phenomena.

Notably, Telegram and Discord nodes exchange invite links almost exclusively with nodes belonging to the same platform. Focusing on the edge counts, Discord-Discord edges (40,040) account for 64.3\% of the total, followed by Telegram-Telegram (11,098, 17.8\%) and Twitter-Telegram (8,371, 13.4\%) edges. Twitter-Discord edges (1,686) are 2.7\% of the total, while Discord-Telegram (527) and Telegram-Discord (556) edges are less than 1\%. According to this results, Discord emerges as a highly interconnected environment, where each channel exchanges invite links with many others. When also taking into account edge weights -- that is, when accounting for the actual number of invite links -- the results are very different. Telegram-Telegram (186,903) edges are 60.1\% of the total, whereas Discord-Discord (60,972) ones end up in second place (19.6\%). Discord-Telegram and Telegram-Discord edges are still very few (less than 1\%). Twitter exhibits strong relationship with Telegram (57,377, 18.5\%), but the amount of invites toward Discord is now very little (less than 1\%). As opposite to Discord, Telegram channels are connected with fewer other channels, but with much stronger links.

In the next section, we deepen our analysis of the invite network by enriching the nodes with semantic features, with the goal of providing explanations for its peculiar structures.

\section{Enriching the invite link network}
\label{sec:enriching}

Tracking cryptocurrency manipulation schemes within the multiform ecosystem enclosed in our dataset requires to overlap layers of knowledge over the map sketched until now. In particular, we want to characterize the nodes of our network according to their genuineness and content traits, in order to drive further analyses.

\subsection{Assessing the nature of Twitter accounts}
\label{subsec:bot}

Besides human users, the Twitter platform is populated by bots. These are accounts controlled by computer algorithms, able to automatically produce content and interact with other accounts, emulating human behaviour. Some bots perform neutral or even useful tasks, but some others instead attempt to manipulate and deceive genuine social media users, pursuing malevolent purposes~\cite{ferrara2016rise}. Hence, in this section we aim at measuring the contribution of Twitter social bots to the diffusion of invite links, in order to evaluate their role in cryptocurrency manipulation schemes.

We performed bot detection on the 7,441 Twitter accounts broadcasting invite links. We used Botometer~\cite{davis2016botornot}, a state-of-the-art Twitter bot detection service, publicly-available via REST API.\footnote{\url{https://rapidapi.com/OSoMe/api/botometer}} Botometer is a supervised machine learning classification model, combining more than a thousand features extracted from profile metadata, friends, social network structure, temporal activity patterns, language and sentiment. The service takes an account ID as input and returns two scores: one is called ``universal'', because it disregards language and sentiment features; the other is specific for English accounts. Since our dataset includes several non-English accounts, we used the universal score, labeling as bots those accounts having a score $\geq 0.5$. 

\begin{figure}[t]
    \centering
    \includegraphics[width=0.8\linewidth, angle=0]{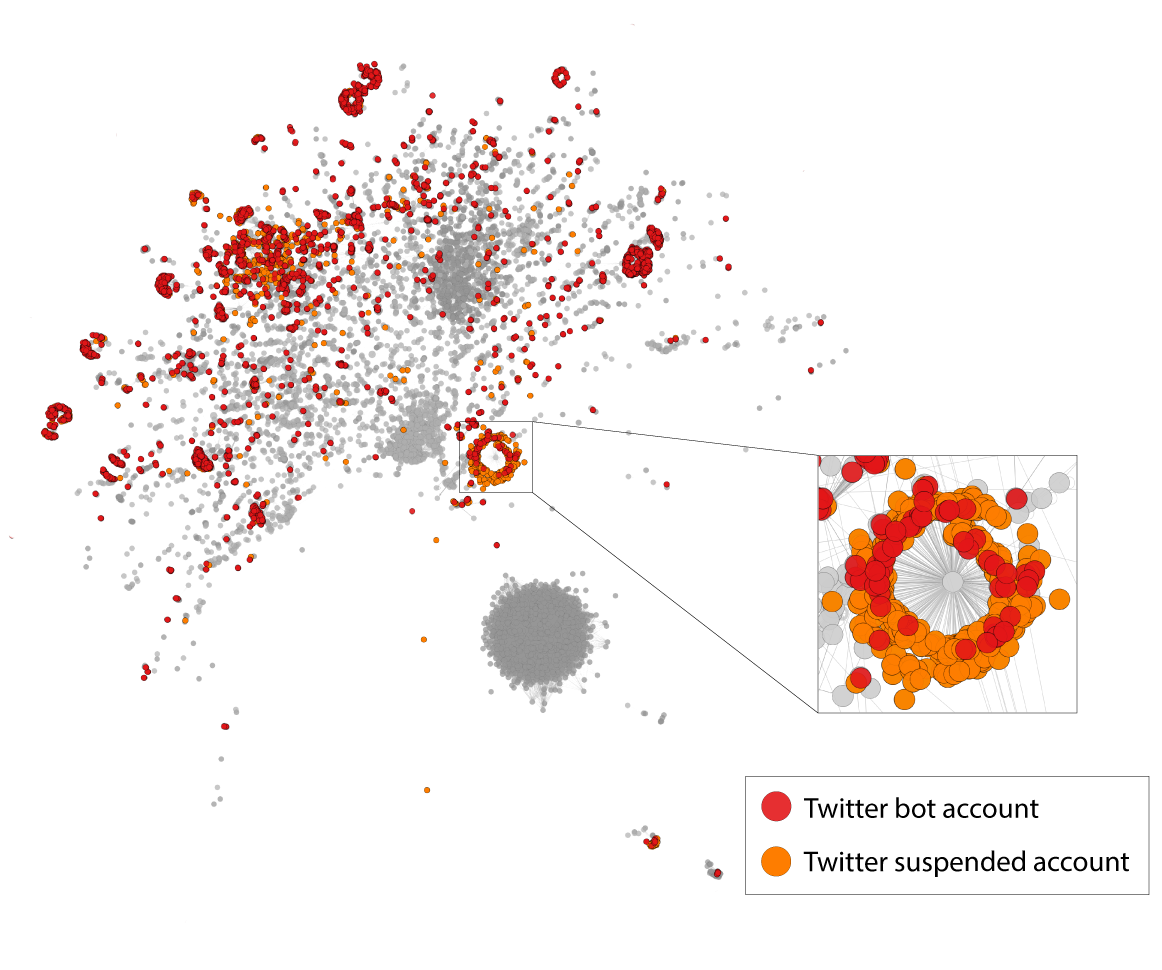}
    \caption{Invite link network highlighting deceptive Twitter accounts. A large portion of Twitter accounts has a deceptive nature (56.3\%). The typical star structures frequently correspond to botnets promoting a single channel. We found 69 botnets with a size of at least 10 elements.
    }
    \label{fig:network-bot}
\end{figure}

Botometer classified 2,710 accounts as bots, resulting in a remarkable fraction of 36.4\% of the total. Other 1,483 (19.9\%) were already suspended by Twitter, again testifying some sort of malicious behavior. By grouping together bots and suspended accounts, we discover that more than a half (56.3\%) of the Twitter accounts involved in invite link broadcasting have a deceptive nature. This remarkable fraction of deceptive accounts largely exceeds previous estimations of overall Twitter bot population, ranging from 9\% to 15\%~\cite{varol2017online}. Instead, it approaches the fraction of 71\% of bots, recently observed when considering most active accounts broadcasting stock-related messages~\cite{cresci2019cashtag}. To this regard, our finding reinforces the knowledge that social bots proliferate in those scenarios involving strong economical incentives. In the remainder, we address both bots and suspended accounts as ``deceptive'' or bots, whereas we define groups of such accounts as ``botnets''.

In Figure~\ref{fig:network-bot}, we highlight Twitter deceptive accounts in the invite link network by coloring the corresponding nodes. Clusters of deceptive accounts clearly emerge, frequently assuming the star shape mentioned in the previous section. At that time, we counted 135 of them in the network, having a size of at least 10 accounts. Now, we find 69 botnets with the same minimum size, 56 of which (81.1\%) are promoting a single Telegram channel. Hence, those star structures can be confidently interpreted as Twitter botnets.

\subsection{Characterizing Discord and Telegram discussions}
\label{subsec:topic}

\begin{table}[t]
\centering
\scriptsize
\begin{tabular}{@{}clr@{}}
\toprule
\textbf{rank}     & \multicolumn{1}{c}{\textbf{words}}                                                                                                     & \multicolumn{1}{r}{\textbf{label}} \\
\midrule
\multicolumn{3}{l}{\textit{Telegram}} \\
1                            & \begin{tabular}[c]{@{}l@{}}\textbf{ref}, \textbf{referral}, \textbf{withdraw}, \textbf{bonus}, paying, \\ doubler, instant, legit, doge, automatic\end{tabular}             & Ponzi scheme                    \\
3                            & \begin{tabular}[c]{@{}l@{}}\textbf{pump}, \textbf{target}, \textbf{signal}, \textbf{stoploss}, market,\\ dump, chart, price, resistance, sell\end{tabular}                 & pump-and-dump                   \\
4                            & \begin{tabular}[c]{@{}l@{}}\textbf{exchange}, \textbf{token}, \textbf{coin}, \textbf{crypto}, \textbf{blockchain}, \\ \textbf{wallet}, \textbf{cryptocurrency}, tokens, exchanges, listed\end{tabular} & legitimate crypto               \\
11                           & \begin{tabular}[c]{@{}l@{}}game, games, fun, players, play,\\ items, item, multiverse, edition, dragons\end{tabular}                   & gaming and entertainment          \\
\midrule
\multicolumn{3}{l}{\textit{Discord}} \\
1                            & \begin{tabular}[c]{@{}l@{}}anime, chill, roblox, nsfw, memes,\\ hangout, gamers, giveaways, chats, fortnite\end{tabular}               & gaming and entertainment          \\
2                            & \begin{tabular}[c]{@{}l@{}}\textbf{wallet}, \textbf{coin}, \textbf{exchange}, \textbf{crypto}, \textbf{blockchain},\\ \textbf{token}, \textbf{cryptocurrency}, btc, coins, address\end{tabular}       & legitimate crypto               \\
9                            & \begin{tabular}[c]{@{}l@{}}\textbf{pump}, \textbf{signal}, \textbf{target}, people, money, \\ high, big, pretty, better, buy\end{tabular}                          & pump-and-dump                   \\
11                           & \begin{tabular}[c]{@{}l@{}}\textbf{referral}, \textbf{bonus}, \textbf{ref}, hosting, \textbf{withdraw}, \\ services, service, network, website, opportunity\end{tabular}    & Ponzi scheme                    \\ \bottomrule
\end{tabular}
\caption{Topic modeling results, obtained by applying Anchored Correlation Explanation (CorEx) to Telegram and Discord channels. Two online cryptocurrency manipulation schemes emerge: Ponzi and pump-and-dump.}
\label{tab:anchored_corex}
\end{table}

In the previous section, we characterized Twitter nodes according to their genuine or deceptive nature. Now, we focus on Telegram and Discord channels, and on the content of the messages shared therein. In detail, we highlight the main topics of discussion within each platform by applying topic modeling. Then, we refine the granularity of our analysis by labeling each channel according to its dominant topic. Notably, a similar approach was already applied to online forums by~\cite{linton2017dynamic}, with interesting results.

To perform topic modeling, we adopted a recent, cutting-edge algorithm known as Anchored Correlation Explanation (CorEx)~\cite{gallagher2017anchored}. As opposed to generative models -- such as Latent Dirichlet Allocation (LDA) -- CorEx learns latent topics over a collection of documents without assuming any particular data generating model. Instead, it leverages the dependencies of words in documents through latent topics, by maximizing the total correlation between groups of words and the respective topic. This approach ensures greater flexibility, enabling hierarchical and semi-supervised variants~\cite{gallagher2017anchored}. In particular, it features word anchoring, a semi-supervised technique improving topic separability with minimum human intervention. In fact, by providing some sets of anchor words relevant for specific topics, it is possible to push the model to better identify and separate them. 

\begin{figure}[t!]
    \centering
    \begin{subfigure}[b]{0.48\linewidth}
        \centering
        \includegraphics[width=\textwidth, angle=0]{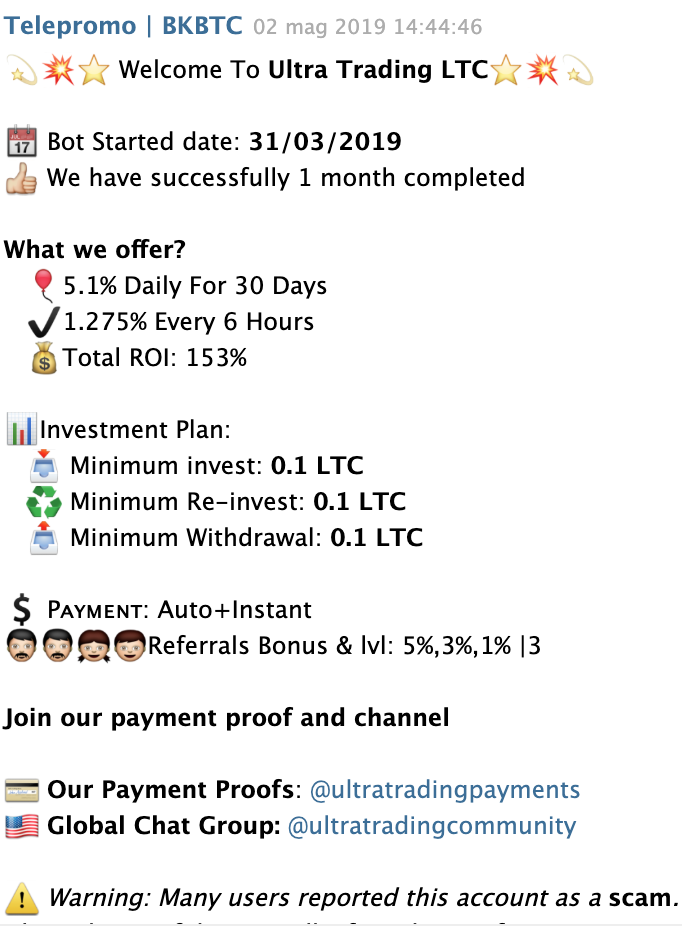}
        \caption{Example of Telegram ``Ponzi scheme'' chat.}
        \label{fig:ponzi_chat_example}
    \end{subfigure}\hspace{0.01\textwidth}
    \begin{subfigure}[b]{0.48\linewidth}
        \centering
        \includegraphics[width=\textwidth, angle=0]{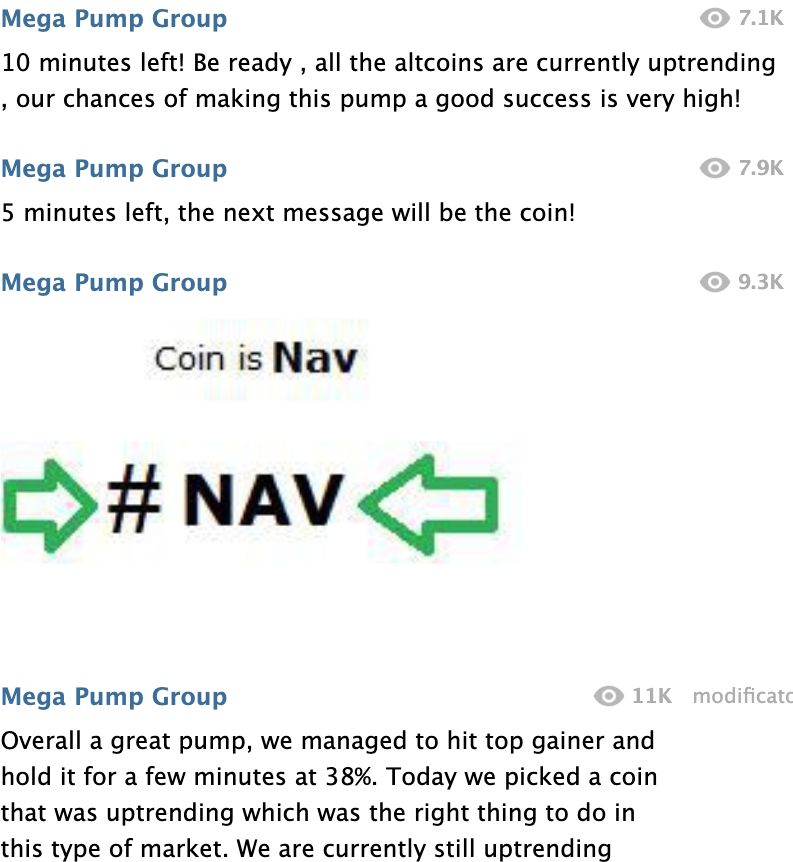}
        \caption{Example of Telegram ``pump-and-dump'' chat.}
        \label{fig:pnd_chat_example}
    \end{subfigure}
    \caption{
     Chats of the cryptocurrency manipulation channels, showing the typical deception patterns outlined as red flags for recognizing Ponzi and pump-and-dump schemes.
    }
    \label{fig:chat_examples}
\end{figure}

\noindent\textbf{Unsupervised topic extraction.} We first applied CorEx without anchoring (i.e., in a completely unsupervised fashion), in order to discover topics spontaneously emerging from our data. To increase the accuracy of our results, we learned two separate models for Discord and Telegram, in order to account for possible differences in topics and forms of speech between the two platforms. In addition, we also filtered channels based on the prevalent language of their messages. In particular, we used the Python library \emph{polyglot}~\cite{alrfouetal2013polyglot} to estimate the prevalent language of each channel, and we neglected non-English ones. As a result, we retained 64.6\% of all Telegram channels and 89.5\% of all Discord channels. In this way, we obtained much more accurate results in terms of detected topics, at the cost of discarding just a minority of all channels involved in cryptocurrency discussions. After experimenting with different configurations, we set the expected number of topics to 12, since additional topics were adding negligible correlation to the learned models. Finally, we ranked the obtained topics according to the fraction of the total correlation that they explain.

For Discord, the topic related to ``gaming and entertainment'' (characterized by words like \emph{anime}, \emph{memes}, \emph{gamers}) explains most of the total correlation of the model. In fourth position, we find a topic related to cryptocurrencies, characterized by generic words like \emph{wallet}, \emph{coin}, \emph{exchange}, \emph{btc}. Regarding Telegram, the most important topic learned by CorEx is characterized by words such as \emph{referral}, \emph{withdraw} and \emph{bonus}. By leveraging results of previous studies~\cite{chen2018detecting}, we are able to connect this topic to the well-known financial scam called \textit{Ponzi scheme}, previously described in the Introduction. As a further confirmation of our labeling, the Telegram messages belonging to this topic share all the features outlined by the U.S. Securities and Exchange Commission~\cite{sec2013ponzi} as red flags for recognizing Ponzi schemes: 
\begin{enumerate*}[label=(\roman*)]
  \item promises of high investment returns with little or no risk,
  \item overly consistent returns,
  \item unregistered investments,
  \item unlicensed seller,
  \item secretive and/or complex investment strategies, and 
  \item no minimum investor qualifications.
\end{enumerate*}
Another interesting finding is that channels associated to this topic are characterized by many similar messages repeatedly posted by Telegram bots, as shown in Figure~\ref{fig:ponzi_chat_example}.
The fourth topic is characterized by words like \emph{pump}, \emph{buy}, \emph{sell} and \emph{resistance}, that can be easily related to \textit{pump-and-dump schemes}~\cite{xu2019anatomy}, previously discussed in the Introduction. Figure~\ref{fig:pnd_chat_example} provides a typical example of a chat in which organizers mobilize participants for the upcoming pump signal. They provide the target coin (e.g., \$NAV) at the scheduled time, and they subsequently comment the results of the operation. In sixth, seventh and ninth positions, we find topics related to legitimate cryptocurrency discussions. One topic includes words related to technological aspects (\emph{blockchain}, \emph{technology}, \emph{platform}), the other two are oriented to finance (\emph{trading}, \emph{investment}). Similarly to Discord, also Telegram has a ``gaming and entertainment'' topic, occurring in the twelfth position.

\begin{figure}[t]
    \centering
    \includegraphics[width=0.8\linewidth, angle=0]{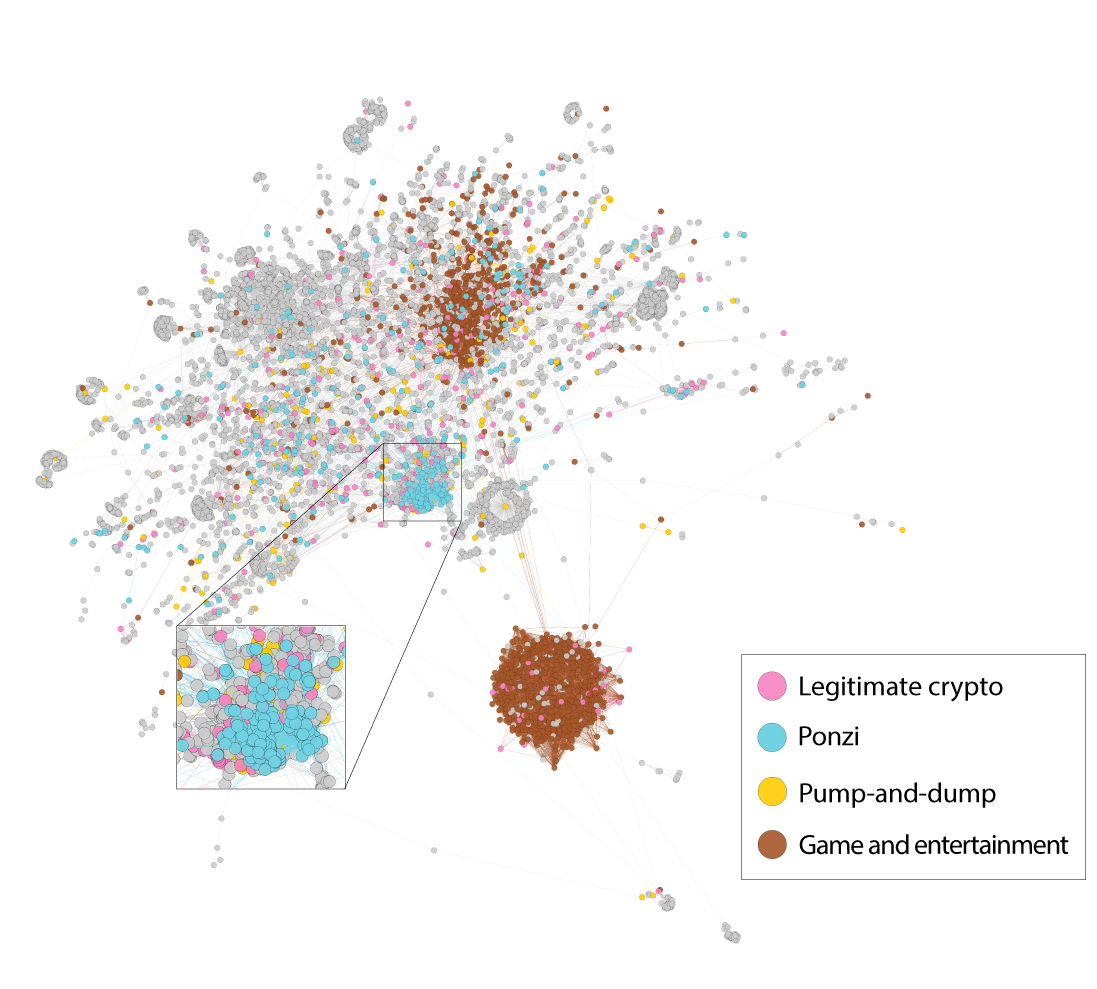}
    \caption{Invite link network with nodes colored according to their prevalent topic. It shows a dense cluster of Ponzi scheme channels engaged in mutual promotion. Instead, pump-and-dump channels are scattered across the network. The weakly connected Discord community is mainly engaged in game and entertainment.}
    \label{fig:network_topics}
\end{figure}

\noindent\textbf{Semi-supervised topic extraction.} Since we are interested in studying manipulations within the cryptocurrency ecosystem, we also leveraged the word anchoring feature of CorEx to improve topic separability, focusing on legitimate cryptocurrency, Ponzi scheme and pump-and-dump topics. We leveraged previous findings, obtained with the unsupervised approach, and domain knowledge derived from existing literature to choose appropriate anchor words. Despite Ponzi scheme and pump-and-dump don't emerge spontaneously on Discord, we leveraged the capability of anchored topic modeling to find underrepresented topics, by forcing the same anchor words as Telegram. The results of this analysis are resumed in Table~\ref{tab:anchored_corex}, where topics are ranked according to the amount of total correlation explained. For each topic, words are ordered according to mutual information with the topic, and anchors are highlighted in bold. Discord is still dominated by the ``gaming and entertainment'' topic. Thanks to anchoring, the legitimate cryptocurrency topic jumped to the second position and improved its quality, as confirmed by the coherence of non-anchored words. Despite anchoring, pump-and-dump and Ponzi schemes confirmed low contribution to correlation and poor internal coherence, showing marginal diffusion among Discord channels. For Telegram, anchoring increased the contribution of our topics of interest to the model correlation. Excellent topic quality was confirmed by the occurrence of non-anchored words with high coherence within each topic, like \emph{dump} and \emph{resistance} for pump-and-dump, or \emph{doubler} and \emph{instant} for Ponzi schemes.

\begin{figure}[t!]
    \centering
    \includegraphics[height=0.2\textwidth, angle=0]{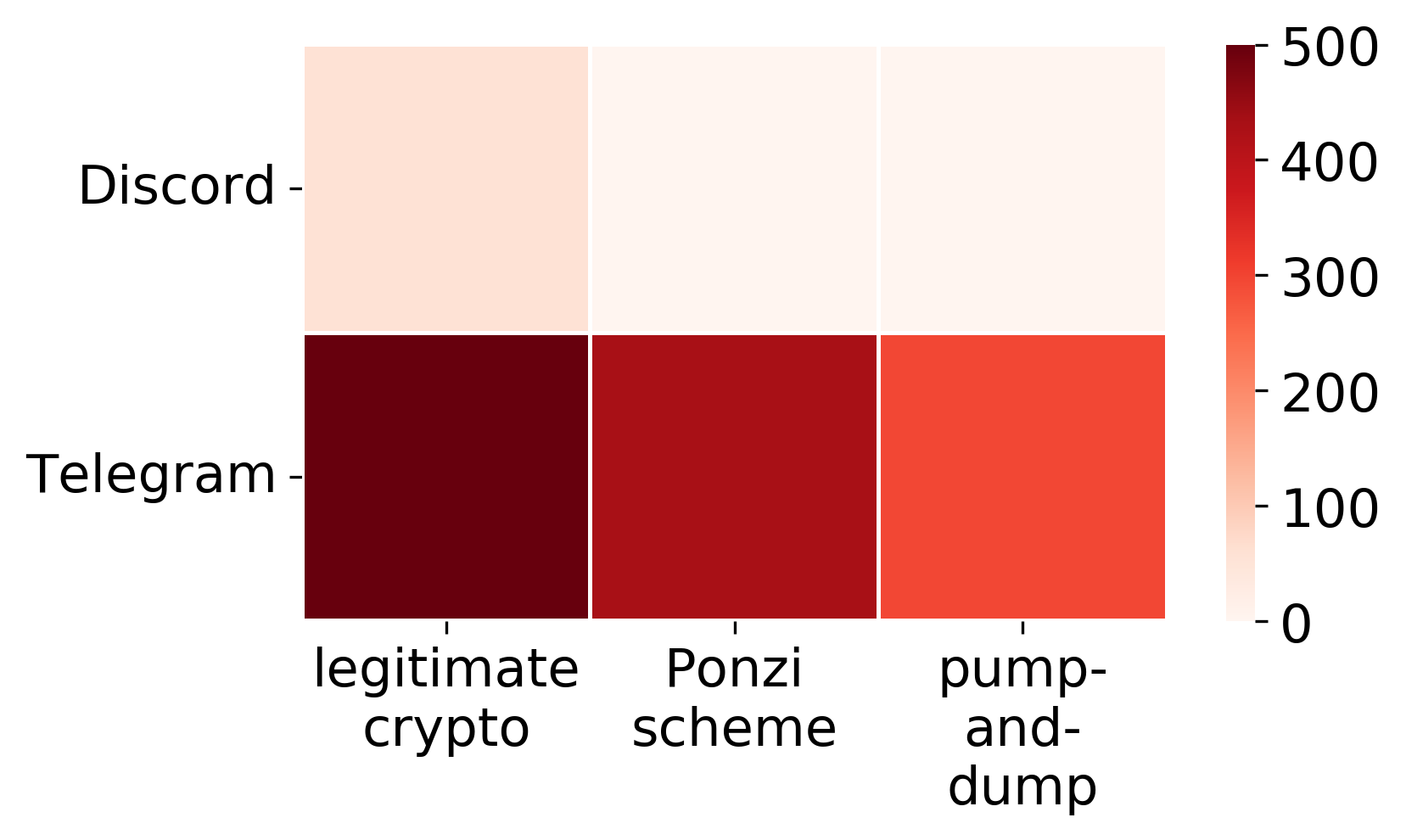}
    \caption{Heatmap of the count of channels per topic by platform. 
    As opposite to Discord, Telegram shows high correlation with cryptocurrency-related topics, together with a remarkable presence of deceptive channels.}
    \label{fig:heatmap_topics}
\end{figure}

\noindent\textbf{Channel labeling.} We used the two semi-supervised models to label Discord and Telegram channels according to their prevalent topic. In particular, we are interested in topics related to legitimate or deceptive cryptocurrency discussion. We leveraged CorEx to compute the correlation of channel textual contents with each possible topic. Then, we labeled each channel with the most correlated topic. Notably, the incidental mentioning of just a few words of a topic in a channel is not sufficient to assign that topic as the channel label. Conversely, prevalent topics are determined by the systematic co-occurrence of the related words. Therefore, this labeling technique ensures sound results, as confirmed by manual inspection. Despite its accuracy cannot reach the one of a supervised classification model, our technique prevents possible biases towards specific cryptocurrency deception schemes that may be introduced by human annotators. In particular, Ponzi and pump-and-dump schemes spontaneously emerged from the data, whereas other well-known schemes (e.g., cryptocurrency thefts) did not. 

In Figure~\ref{fig:network_topics}, we color the invite link network nodes according to the assigned topic. Uncolored nodes correspond to Twitter accounts, non-English channels and channels with a non-labeled (i.e., generic, uninteresting) topic. The isolated community of Discord channels, already mentioned in the previous section, is clearly dominated by the gaming and entertainment topic (86.2\% of nodes). A cluster of Ponzi scheme Telegram channels clearly emerges, which approximately corresponds to the cluster of Telegram channels magnified in Figure \ref{fig:network-graph}. Conversely, pump-and-dump and legitimate cryptocurrency channels are scattered across the network. In Figure~\ref{fig:heatmap_topics}, a heatmap shows the channel counts per topic per platform, focusing on cryptocurrency-related topics. Consistently with topic modeling results, Discord has low correlation with cryptocurrencies, with 58 channels labeled as legitimate cryptocurrency (3.3\%), only one pump-and-dump and zero Ponzi scheme. On the contrary, Telegram hosts 504 legitimate cryptocurrency (13.2\%), 432 Ponzi scheme (11.3\%), and 296 pump-and-dump (7.8\%) channels. Hence, the high correlation with cryptocurrency-related topics is confirmed, together with a remarkable presence of cryptocurrency manipulation channels.

\section{Uncovering cryptocurrency manipulations}\label{sec:manipulation}

\begin{figure*}[ht!]
    \centering
    \begin{subfigure}[b]{.47\textwidth}
    \centering
        \includegraphics[height=0.55\textwidth, angle=0]{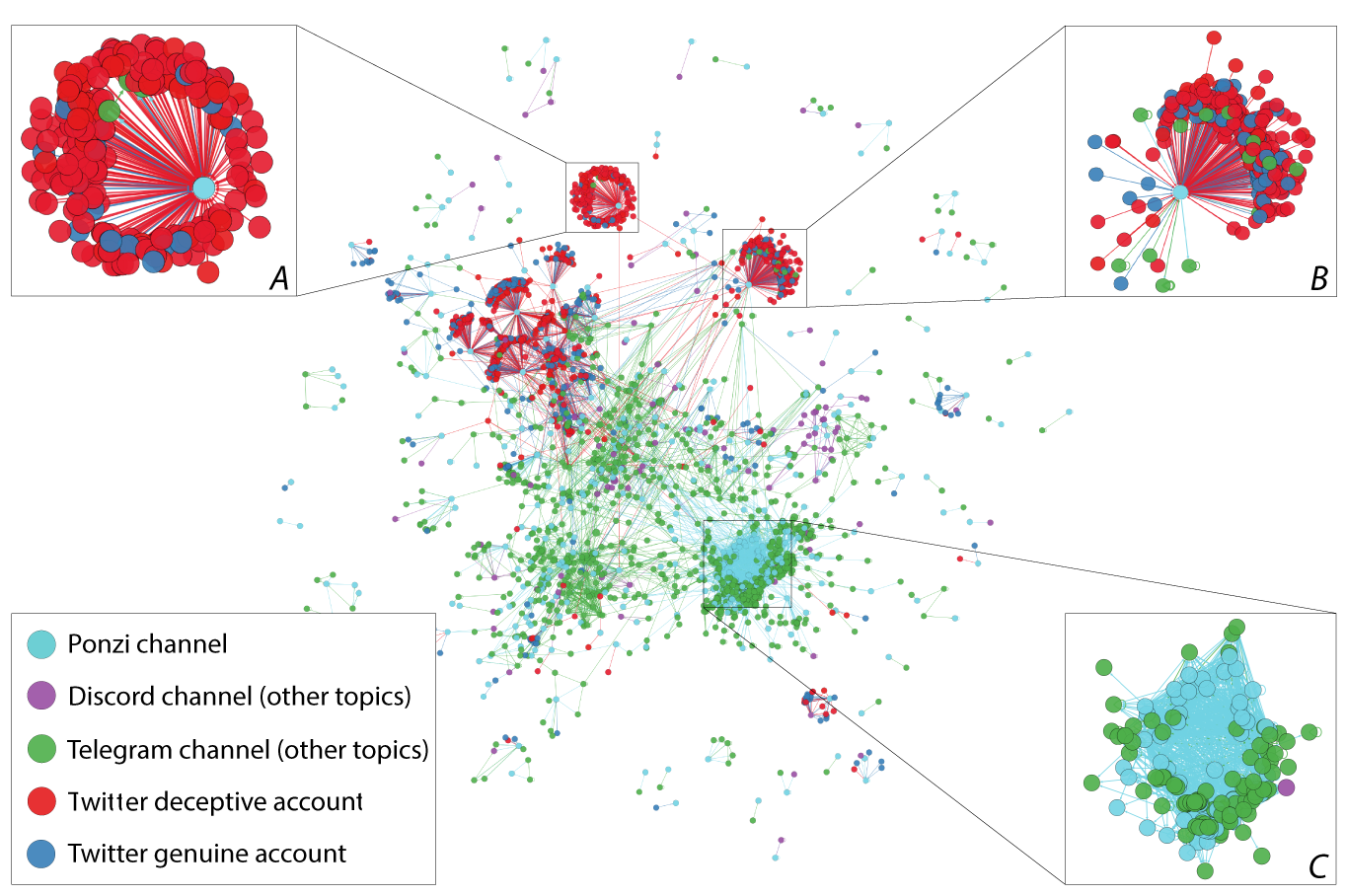}
         \caption{Neighbour network of Ponzi scheme channels.}
        \label{fig:network-ponzi-ego}
    \end{subfigure}
    \begin{subfigure}[b]{.47\textwidth}
    \centering
        \includegraphics[height=0.55\textwidth, angle=0]{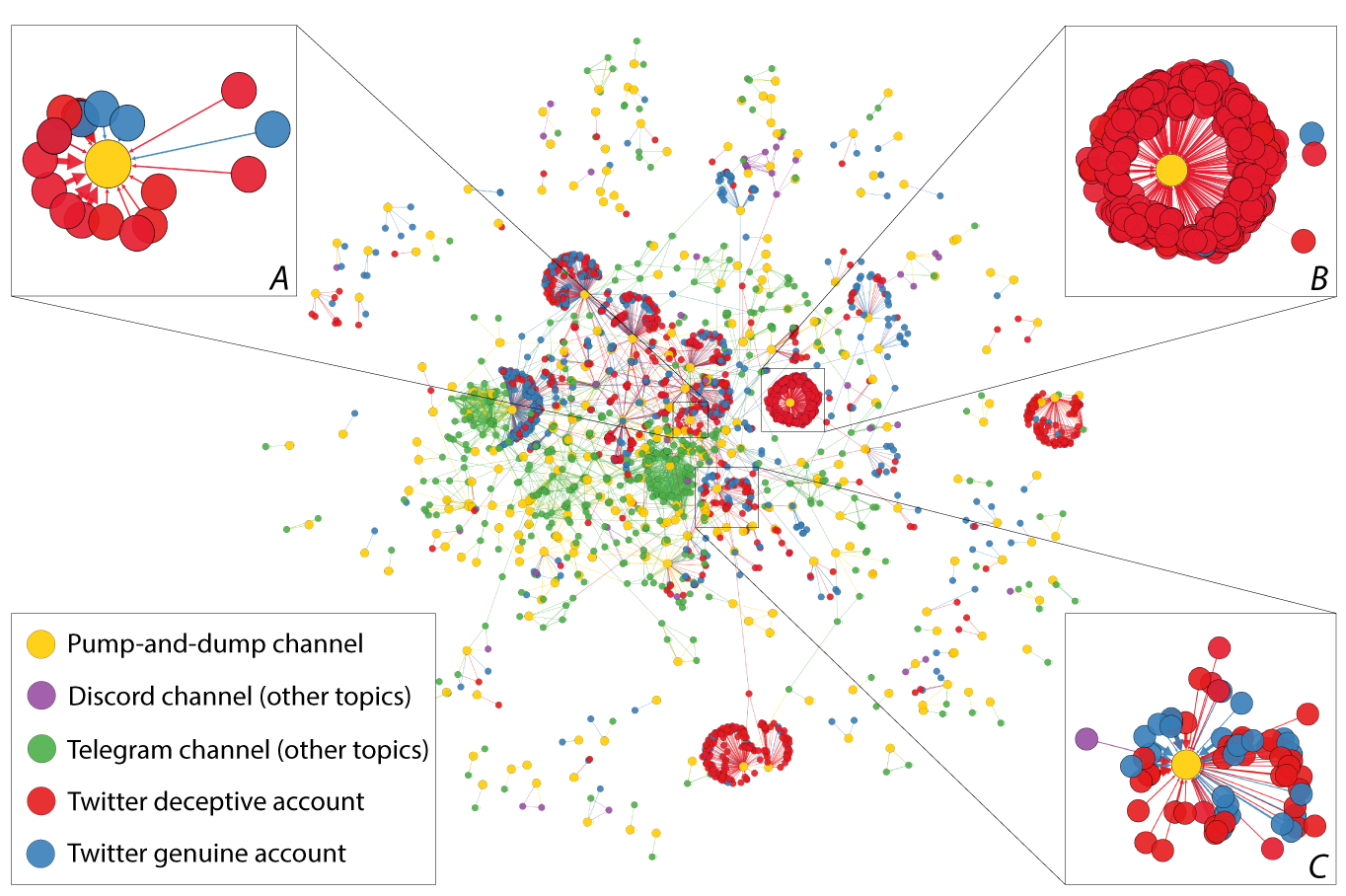}
        \caption{Neighbour network of pump-and-dump channels.}
         \label{fig:network-pnd-ego}
    \end{subfigure}
    \caption{
    Portions of the invite link network in direct contact with deceptive Ponzi scheme (\ref{fig:network-ponzi-ego}) and pump-and-dump channels (\ref{fig:network-pnd-ego}).
    While Ponzi scheme channels are strongly engaged in mutual promotion, pump-and-dump channels are mainly endorsed by star structured Twitter botnets.}
    \label{fig:ego_networks}
\end{figure*}

In previous sections, we sketched a map of the online cryptocurrency landscape, by building the invite link network. 
Then, we added two semantic layers. The first one allowed us to label Twitter nodes according to their genuine or deceptive nature, while the second one characterized Telegram and Discord channels according to their prevalent topic of discussion. In this way, two schemes of deception naturally emerged: pump-and-dump and Ponzi scheme. Now that we have charted the online landscape of cryptocurrency manipulations, we leverage our map for investigating the tracks of manipulation. Firstly, we ``zoom in'' to focus on the portions of the original network in direct contact with the deceptive channels. Then, we ``zoom out'' to interpret our results within the general framework of online manipulation.

\begin{figure}[t]
    \centering
    \begin{subfigure}[t]{.46\linewidth}
        \includegraphics[width=\linewidth]{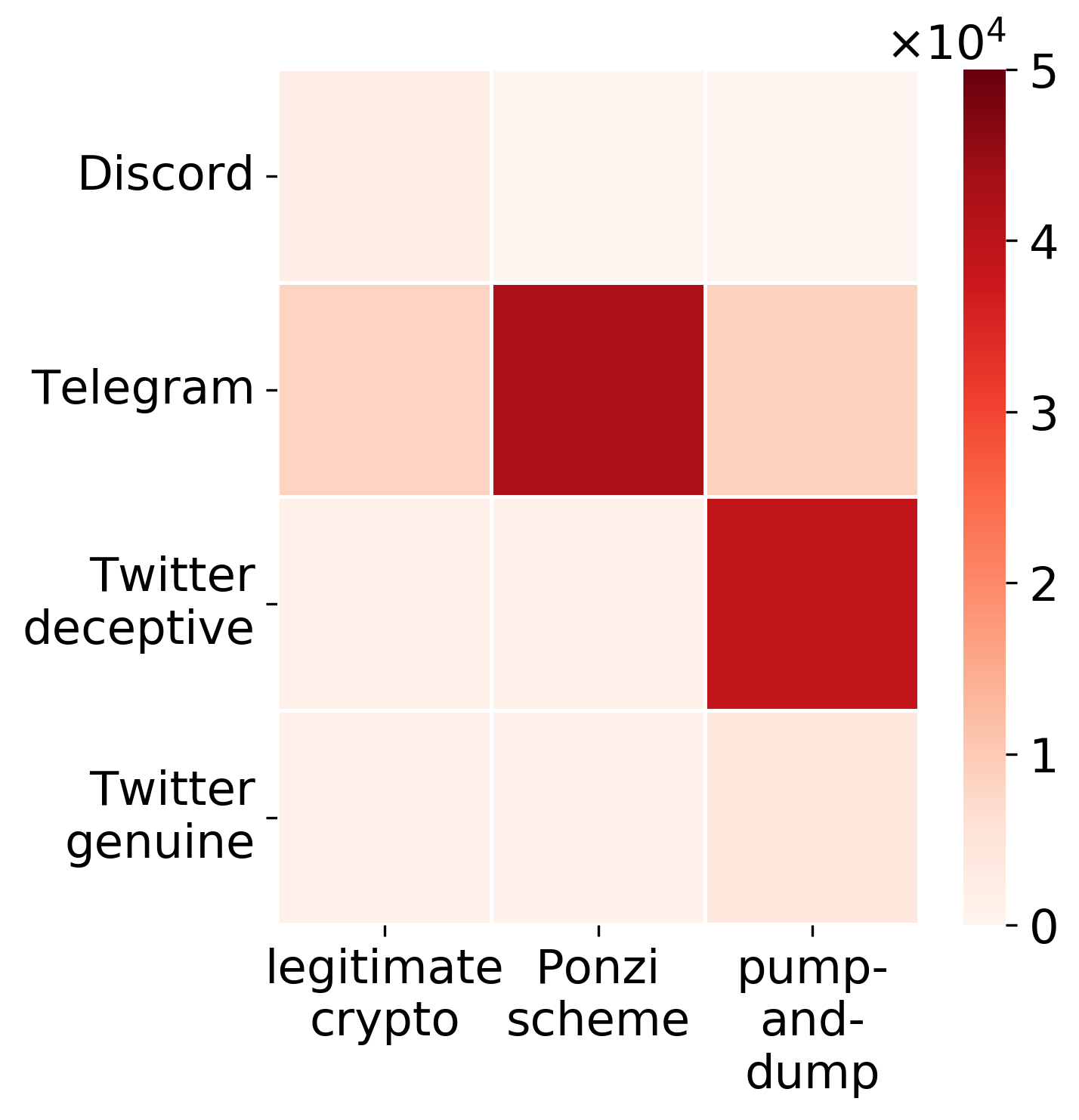}
        \caption{Heatmap of the number of invites per source node platform and target node topic. }
        \label{fig:source_target_bot}
    \end{subfigure}\hspace{0.01\textwidth}%
    \begin{subfigure}[t]{.51\linewidth}
        \includegraphics[width=\linewidth]{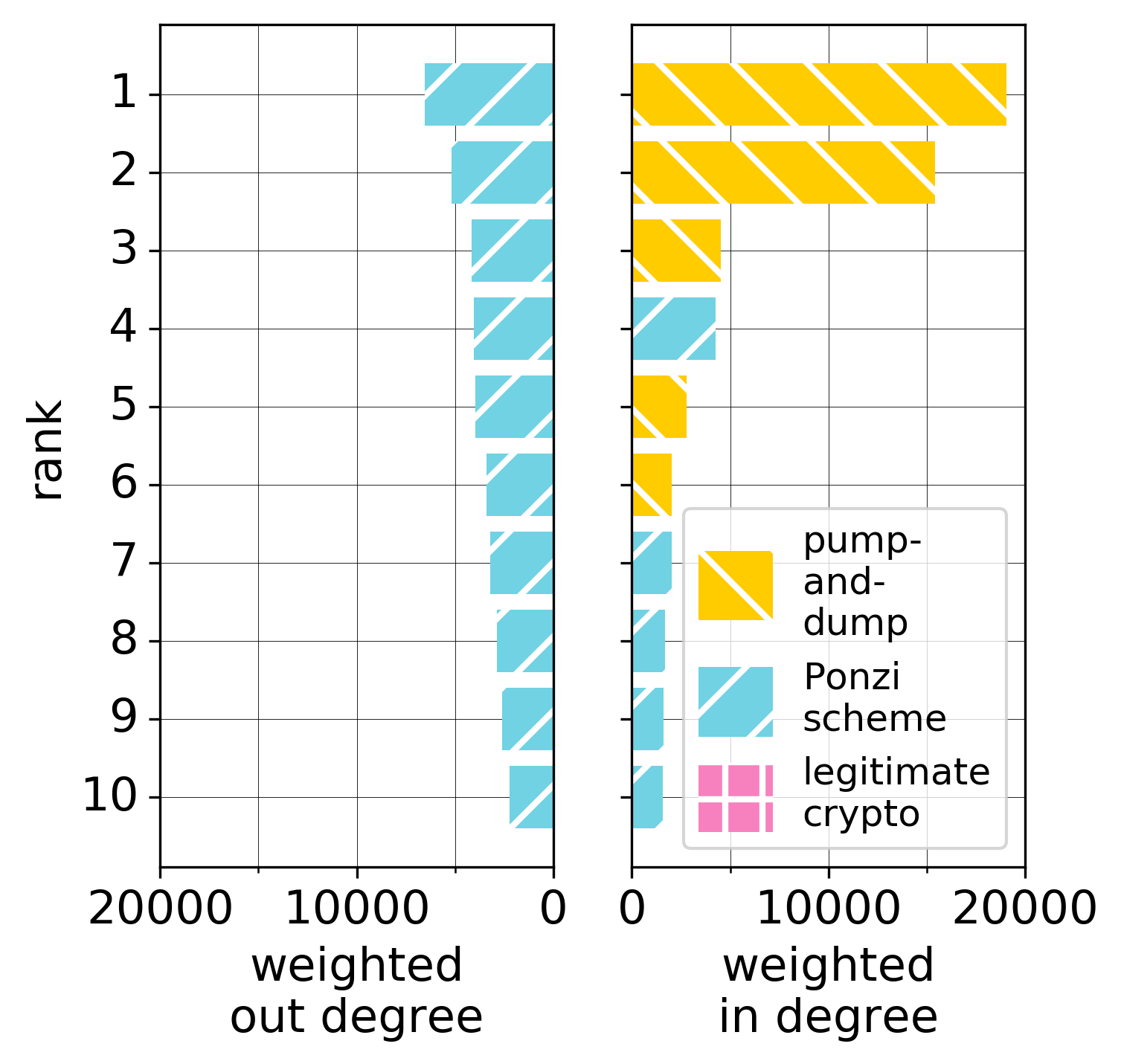}
        \caption{Top-10 cryptocurrency-related channels, sorted by weighted out (left) and in (right) degrees. Bars are colored according to the channel prevalent topic.}
        \label{fig:topic_degree_ranking}
    \end{subfigure}\hspace{0.03\textwidth}%
    \caption{Interplay between invite link diffusion and channel behaviour. Figure~\ref{fig:source_target_bot} shows that deceptive channels attract the most part of invites (87.8\%). Ponzi scheme channels are mainly promoted within Telegram, while pump-and-dump channels receive most of the invites from Twitter deceptive accounts.
    Figure~\ref{fig:topic_degree_ranking} confirms deceptive channels as the major hubs of invite diffusion. In fact, no legitimate cryptocurrency channel ranks in the top-10 in/out degree nodes.}
\end{figure}

\noindent\textbf{Ponzi schemes.} In Figure~\ref{fig:network-ponzi-ego}, we isolate Ponzi scheme nodes, their first neighbours and the related edges. The 432 Ponzi scheme channels are colored in pale blue, whereas other channels are colored according to the platform they belong to. We distinguish Twitter accounts according to their genuine (blue-colored) or deceptive (red-colored) nature. The scene is dominated by Telegram and Twitter platforms (1,124 and 1,696 nodes, respectively), with only 106 Discord nodes. We count 600 genuine and 1,096 deceptive Twitter accounts, resulting in a fraction of deceptive accounts of 64.6\%, significantly higher than the one measured for the whole network (56.3\%). There are 11 Twitter botnets having a size of at least 10 nodes, promoting a Ponzi scheme channel. They account for the 15.9\% of the total, while Ponzi scheme channels are only the 7.8\% of the total amount of Discord and Telegram channels. Two examples of those botnets are magnified in Figure~\ref{fig:network-ponzi-ego}, in panels \textit{A} and \textit{B}. As shown, they feature the typical star structure that we previously highlighted. In panel \textit{C}, we also highlight a dense cluster of Ponzi scheme Telegram channels, roughly corresponding to the one shown in Figure~\ref{fig:network_topics}. This cluster is the largest cryptocurrency manipulation hub found in our study. It is composed of 166 Telegram nodes, 63 (39.8\%) of which are Ponzi scheme channels. To understand its role within the Ponzi scheme ecosystem, in Figure~\ref{fig:source_target_bot} we represent the heatmap of the number of invites per source node platform and target node topic. For source nodes, we also separate genuine Twitter accounts from deceptive ones. Results show that Ponzi scheme channels collect 71.4\% of invite links shared by Telegram source nodes in the whole network. Moreover, in 92.3\% of cases, invites targeting Ponzi scheme channels originated from other Ponzi scheme channels. Hence, most of the diffusion of invites to Ponzi scheme channels was carried out, within the examined cluster, by other Ponzi scheme channels. The engagement on mutual promotion within the Ponzi scheme cluster is further confirmed by Figure~\ref{fig:topic_degree_ranking}, showing that the top-10 channels with highest weighted out degree perform Ponzi schemes. 

\noindent\textbf{Pump-and-dump.} In Figure~\ref{fig:network-pnd-ego}, we depict the neighbour network of pump-and-dump channels. Pump-and-dump channels are colored in yellow, whereas for other nodes we apply the same convention as before. Besides the 297 pump-and-dump nodes, the network is composed of 1,917 Twitter, 504 Telegram and 52 Discord nodes. Pump-and-dump nodes are scattered across the network, and it is not possible to identify any cluster of them. Also in this case, the fraction of Twitter accounts having a deceptive nature (65.4\%) significantly exceeds the one measured on the whole network. They are frequently organized in botnets. In detail, we spot 15 botnets with a size of at least 10 accounts, promoting a pump-and-dump channel. They account for the 21.7\% of the observed botnets, resulting overrepresented if we consider that pump-and-dump are only 5.3\% of the total channels. To estimate the contribution of those botnets in promoting pump-and-dump channels, we again resort to the heatmap of Figure~\ref{fig:source_target_bot}. We find out that Twitter deceptive accounts contribute to the 75.4\% of all the invite links to pump-and-dump channels. Conversely, 92.9\% of invite links, diffused by Twitter deceptive accounts, point to pump-and-dump channels. The effectiveness of Twitter deceptive accounts in promoting pump-and-dump channels is further proved by Figure~\ref{fig:topic_degree_ranking}, showing that five of the top-6 channels with highest weighted in degree are labeled as pump-and-dump. The first three of them are magnified in Figure~\ref{fig:network-pnd-ego}. They appear surrounded by their respective botnets, responsible for the high weighted in degree of their target channels. The botnet in panel \textit{B} promotes the MET$\Delta$.Symetra Telegram channel, resulting in the star structure that was magnified in figures \ref{fig:network-graph} and \ref{fig:network-bot}. This channel and its botnet represent the largest 
invite diffusion hub in our study. Yet, to the best of our knowledge, the existence of this pump-and-dump channel was unknown prior to our analysis, since it was never mentioned in existing studies, nor it is reported in authoritative lists of known pump-and-dump channels~\cite{xu2019anatomy}. This striking result further supports the soundness of our method and the impact of our findings.

\noindent\textbf{Zooming out to the general framework.} Our exploration of the online cryptocurrency ecosystem confirms the concerns about the susceptibility of cryptocurrency markets to online manipulation, raised by authoritative agencies~\cite{sec2018ponzi,sec2013ponzi}. While Discord appears as an overall healthy environment, Twitter and Telegram reveal a strong interplay between numerous deceptive agents, engaged in promoting scams. The choice of the invite link diffusion as the compass orienting our route proved to be particularly suitable for tracking online cryptocurrency manipulations. It was motivated by two hypotheses:
\begin{enumerate*}[label=(\roman*)]
  \item the exchange of invite links implies homophily and common goals between the involved agents, and
  \item cryptocurrency manipulation stimulates invite link diffusion, because the efficacy of deceptive schemes strongly depends on recruiting a large number of participants. 
\end{enumerate*}
The first claim is supported by the existence of the dense cluster of Telegram Ponzi scheme channels, strongly committed in mutual promotion. Further confirmation comes from the finding of several Twitter botnets, specially created to promote pump-and-dump channels. In both cases, agents sharing similar features, behaviours and goals result strongly connected by the invite link diffusion. The second intuition is brilliantly confirmed by results shown in Figure~\ref{fig:topic_degree_ranking}, proving that legitimate channels collect a negligible fraction of the overall invite links (12.2\%). In contrast, cryptocurrency manipulation emerges as the main trigger for invite link diffusion in the online cryptocurrency ecosystem. 

Our study allows to estimate the alarming extent of deception in the online cryptocurrency ecosystem. In fact, the 56.5\% of the cryptocurrency-related channels in our dataset is involved in deception, despite the fact that we avoided to bias our data crawling towards them, as instead done in the majority of previous works~\cite{xu2019anatomy,mirtaheri2019identifying}. Moreover, Twitter botnets emerge as the main vehicle for spreading pump-and-dump invites. This result enriches our knowledge on Twitter bot activities with a new element, relating our work with the flourishing line of research that aims to estimate how social bots manage to condition human activities in various ways, from contaminating the social debate~\cite{stella2018bots} to adulterating the economic processes~\cite{cresci2019cashtag,cresci2018fake}. 
Notably, our findings are not merely descriptive, but they provide actionable knowledge to counteract cryptocurrency manipulations. In fact, tracking the major hubs of invite diffusion is a simple, effective way to spot malicious agents and manipulation schemes, as proven by the discovery of the previously unknown deceptive channels. Moreover, the success of these manipulations depend on the possibility to exploit invite link and Twitter bots. Hence, limiting the diffusion of invites and reducing the activity of bots would severely impair the efficacy of these frauds. Enforcing such actions could be particularly relevant for authorities responsible for the safety of the online financial markets, like the U.S. Securities and Exchange Commission and the U.S. Commodity Futures Trading Commission.


Cryptocurrencies were born to empower the dream of an accountable, decentralized, democratic payment method, preserving the user privacy and subtracting the consumer habits to the undesired scrutiny of governments and corporations.
In the same way, the Web was meant to realize the promise of an ecosystem granting free speech and equal access to information, goods and opportunities to every human being. Unfortunately, the conjunction between the potentialities of cryptocurrencies and the Web has opened the Pandora's box of criminal darknet markets, wild financial speculation, money laundering, criminal and terrorist organization financing and deceptive manipulations~\cite{brown2016cryptocurrency}. This work addresses a peculiar example of those threats. Despite its specificity, typical patterns of online deception emerged, confirming the pervasiveness of these nasty phenomena across multiple aspects of online human activities. This work thus contributes towards raising collective awareness about the risks and the opportunities offered by cryptocurrencies to our Society, and to stimulate further research for designing countermeasures to the related threats.

\section{Concluding remarks}\label{sec:conclusion}

Motivated by the increasing alarm raised by institutions about cryptocurrency manipulation, we mapped the online cryptocurrency ecosystem to identify, assess and characterize possible threats. By cross-checking over 50M messages across Twitter, Telegram and Discord platforms, we analysed the diffusion of invite links to cryptocurrency-related channels. Results confirmed our controlling idea, based on the hypothesis that invite link exchange is a proxy for homophily and common goals between the involved agents, as well as a characteristic pattern related to deceptive schemes. First, we observed that two cryptocurrency manipulation schemes emerged -- ``pump-and-dump'' and ``Ponzi'' -- both affecting Telegram much more than Discord. Then, we identified a dense cluster of Ponzi scheme channels, so engaged in mutual promotion as to contribute to the 71.4\% of the overall invite link diffusion measured on Telegram. Finally, we reported on 15 Twitter botnets that are responsible for the 75.4\% of invite links to pump-and-dump channels, thus adding a new piece of knowledge about social bot activities.   

Since institutions are evaluating the eligibility of cryptocurrencies as legal payment method, our research community must raise awareness and design countermeasures to possible threats related to this emerging scenario. 
This work provides actionable knowledge, suitable to enforce more effective responses. As part of our follow up work, we
plan to cross-check our social media data with the cryptocurrency price trends to predict upcoming manipulations. 

\bibliographystyle{aaai}
\fontsize{9.0pt}{10.0pt}
\selectfont
\bibliography{main.bib}

\end{document}